%% file: arxiv.tex
\documentclass[11pt]{article}
\usepackage[utf8]{inputenc}
\title{Private federated  discovery of out-of-vocabulary words for Gboard}
\author{Ziteng Sun, Peter Kairouz, Haicheng Sun, Adria Gascon, Ananda Theertha Suresh}
\date{
\texttt{\{zitengsun, kairouz, haicsun, adriag, theertha\}@google.com} \\[2ex]
Google}
\input{glodef}

\input{locdef}

\usepackage[numbers]{natbib}
\usepackage{subcaption}
\usepackage{graphicx}
\usepackage{fullpage}

\usepackage{xcolor}
\usepackage{bbm}
\renewcommand{\ignore}[1]{}
\usepackage[capitalise]{cleveref}

\newcommand{\bz}{\mathbf{z}}
\newcommand{\bv}{\mathbf{v}}
\newcommand{\bV}{\mathbf{V}}
\def\withcolors{0}
\def\withnotes{0}

\ifnum\withnotes=1
\renewcommand{\th}[1]{ {\noindent \textit{\small\textcolor{orange}{theertha: #1}}}}
\newcommand{\zs}[1]{{\noindent \textit{\small\textcolor{red}{ziteng: #1}}}}
\newcommand{\znote}[1]{ \textcolor{purple}{Ziteng: #1} }
\newcommand{\tnote}[1]{ \textcolor{blue}{Theertha: #1} }
\newcommand{\anote}[1]{ \textcolor{orange}{Adria: #1} }
\newcommand{\todo}[1]{ \textcolor{red}{TODO: #1} }
\else
\renewcommand{\th}[1]{{}}
\newcommand{\zs}[1]{{}}
\newcommand{\znote}[1]{}
\newcommand{\tnote}[1]{}
\newcommand{\anote}[1]{}
\newcommand{\todo}[1]{}
\fi

\ifnum\withcolors=1

\else

\fi

\newcommand{\coloredcomment}[1]{\COMMENT{\textcolor{blue}{#1}}}
\newcommand{\eow}{\bot}
\newcommand{\unk}{\gamma}
\newcommand{\sample}{\textsc{Sample}}
\newcommand{\gs}{\textsc{GreedySampling}}
\newcommand{\rs}{\textsc{RandomSampling}}
\begin{document}

\maketitle
\begin{abstract}
The vocabulary of language models in Gboard, Google's keyboard application, plays a crucial role for improving user experience. One way to improve the vocabulary is to discover frequently typed out-of-vocabulary (OOV) words on user devices. This task requires strong privacy protection due to the sensitive nature of user input data. In this report, we present a private OOV discovery algorithm for Gboard, which builds on recent advances in private federated analytics. The system offers local differential privacy (LDP) guarantees for user contributed words. With anonymous aggregation, the final released result would satisfy central differential privacy guarantees with $\eps = 0.315, \delta = 10^{-10}$ for OOV discovery in en-US (English in United States). %

\end{abstract}

\section{Introduction}

Gboard relies on language models (LMs) to improve user experience via features like next word prediction and autocorrection \cite{gboard-blogpost, ouyang2017mobile, hard2019federated, chen2019federated}. The performance of LMs depends on their vocabulary, a predefined list of words that they can process. For example, words might be corrected wrongly if they are missing from the vocabulary~\cite{ouyang2017mobile}. %
Discovering frequently typed words from users could help engineers and analysts improve the vocabulary, and hence the LMs based on it~\cite{chen2019federated2}. %
However, for keyboard applications like Gboard, the task is particularly challenging due to various reasons: (1) User input can contain highly sensitive information such as their addresses, phone numbers, or even social security numbers. This would require the algorithm to provide stringent privacy protection against user data. (2) The data is from a large open domain\footnote{An open domain is a domain whose partitions (keys) are determined from the dataset itself (i.e. they are not fixed a priori).}, which technically can contain strings of arbitrary length, putting extra requirement on the efficiency of the system.
(3) For OOV discovery tasks, the data is typically more heavy-tailed than the in-vocab words, making it hard for differentially-private discovery algorithms~\cite{zhu2020federated}, which usually requires certain form of randomization such as sampling and noising.

Federated heavy hitter discovery is a 
technique that allows the server to discover frequent input elements in distributed datasets without sharing sensitive data. Differential privacy \cite{dwork2006calibrating}, and its local version \cite{kasiviswanathan2008ldp} could be added on top to further provide formal privacy guarantees on the discovering process.
Heavy hitter discovery with LDP guarantee has been studied both in academia and industry~\cite{ErlingssonPK14,bassily2017practical,apple_privacy}. 
In recent years, there has been a line of works~\cite{zhu2020federated,  bagdasaryan2021towards, cormode2022sample, mcmillan2022private, chadha2023differentially} that focus on the open-domain setting %
and learn heavy hitters with multiple rounds of interaction between the server and user devices using a trie (prefix tree) data structure. They also adopt recent advances in private analytics such as privacy amplification via shuffling~\cite{ErlingssonFMRTT19, feldman2021hiding, feldman2023stronger} to enhance the utility and privacy guarantees of federated heavy hitter algorithms.

Gboard has a record of holding high standards for the use of user data in improving language models \cite{xu2023federated, fl_gboard_blog}. 
In this report, we build upon the aforementioned line of work and present an open-domain federated heavy hitter discovery system that utilizes state-of-the-art algorithms in differential privacy and federated analytics \cite{fa_blog}. %
The algorithm provides LDP guarantee on user contributed words, as well as central differential privacy guarantee with better privacy parameters using privacy amplification analysis~\cite{feldman2021hiding, feldman2023stronger}.  We use the algorithm to perform private OOV discovery with Gboard inputs from en-US (English in United States) users and demonstrate the practicality of the algorithm.

\section{Private federated heavy hitter discovery}
We consider heavy hitter discovery in the federated setting with multiple rounds of communication between the server and a user population. The population has $\ns$ users, and each of them contributes over a period of $T$ days. %
Let $S_i$ be the set of all user contributions. %
Our goal is find frequently contributed elements in the union of these sets $\mathbf{S} = (S_1, S_2, \ldots, S_\ns)$.

In the open domain setting, we consider the case where each contributed element is a string consisting of predefined characters from a small set $\cX$. The set of all possible  user contributions would be 
\[
    \cZ = \cX^* \eqdef \cup_{\ell = 1}^\infty \cX^\ell.
\]

Each element $\bz \in \cZ$ can be represented as a sequence $\bz = z[1]z[2]\ldots z[{\ell}]$ where $\ell = |\bz|$ and $\forall i \in [\ell]$\footnote{We use $[\ell]$ to denote the set $\{1, 2, \ldots, \ell\}$.}, $z[i] \in \cX$. Without loss of generality, we assume $\cX$ contains a predefined \textit{end of sequence} symbol $\eow$ and all elements would end with $z[\ell] = \eow$. For example, for OOV discovery in en-US, $\cX$ contains the list of letters, digits, and special characters. %

\paragraph{Differential privacy.} To provide strong formal privacy protection against user inputs, we consider various notions of differential privacy, which requires that the algorithm output is insensitive to  individual contributions from the users.

\begin{definition}[Central Differential Privacy (DP)~\citep{dwork2006calibrating}] We say an algorithm $
\cA$ is $(\eps, \delta)$-differentially private, if for any pair of neighbouring datasets $\bS \sim \bS'$ (will define this later), and any output $O \subset \cO$, we have
\[
    \probof{\cA(\bS) \in O} \le e^\eps \cdot \probof{\cA(\bS') \in O} + \delta.
\]
\end{definition}

Here the neighbouring relation\footnote{We focus on the swap model of differential privacy in this paper.} depends on the unit of privacy we choose for the application. The privacy units include: (1) Item-level: $\bS \sim_{\rm item} \bS'$ \iff $\bS$ and $\bS'$ differ at a single contribution from a user; 
(2) User-level: $\bS \sim_{\rm user} \bS'$ \iff $\bS$ and $\bS'$ differ at the whole contribution from a user in all days. %

\begin{definition}[Local differential privacy (LDP) \cite{kasiviswanathan2008ldp}] Let $Q$ be a randomized mapping from the an input set to the message space $\cY$. We say $Q$ is an $\eps$-LDP randomizer if it satisfies that $\forall S, S'$ and $y \in \cY$, we have
\[
    \probof{Q(S) = y } \le e^\eps \cdot \probof{Q(S') = y}.
\]
An algorithm $\cA$ is said to be $\eps$-LDP at item-level (user-level) if it satisfies that $\cA$ is a post-processing of messages from $\eps$-LDP randomizers on each contributed item (each user's contribution set).
\end{definition}

\section{Trie-based heavy hitters with local differential privacy}
Next we present the interactive algorithm that allows us to discover heavy hitters in an open domain with LDP guarantee at item-level. Moreover, upon aggregation, the final output of the algorithm will satisfy central DP with stronger privacy parameters.

The algorithm is a variant of the TrieHH algorithms in \cite{zhu2020federated} and is closely related to various interactive algorithms that use tree data structures to learn heavy hitters with DP \cite{bassily2017practical, bagdasaryan2021towards, mcmillan2022private, chadha2023differentially}. To get LDP guarantee on user contributions, we apply an LDP randomizer on user contributions when learning each layer of the trie (prefix tree) structure. For the LDP randomizer, we use the Subset Selection~\citep{wang2016mutual, YeBarg2017optimal} algorithm, presented in \cref{alg:subset_selection}.
\begin{algorithm}[h]
\caption{${\mathbf{SS}}(z, \cZ, \eps).$ Subset Selection~\citep{wang2016mutual, YeBarg2017optimal}.}
\begin{algorithmic}[1]
\STATE \textbf{Input:} List of possible contributions $\cZ$, user input: $z \in \cZ$, LDP parameter $\eps$.
\STATE Set $d = \ceil{\frac{s}{e^\eps + 1}}$, $p = \frac{d e^\eps}{d e^\eps + s - d}$ where $s = |\cZ|$.
\STATE $S = \emptyset$
\STATE Draw a uniform random variable $\zeta \sim \text{Unif}([0,1])$.
\IF{$\zeta < p$}
\STATE Add $z$ to $S$.
\STATE Select $d - 1$ random elements from $\cZ \setminus \{z\}$ and add them to $S$.
\ELSE
\STATE Select $d$ random elements from $\cZ \setminus \{z\}$ and add them to $S$.
\ENDIF
\STATE \textbf{Return} $S$.
\end{algorithmic}
\label{alg:subset_selection}
\end{algorithm}

The algorithm has been shown to achieve the optimal privacy-utility tradeoff for frequency estimation under $\ell_2$ metric. However, for sets with large, or even infinite support, such as the case where $\cZ = \cX^*$ for some $\cX$, the algorithm will be less efficient in terms of communication and computation cost. To resolve this, we build upon the TrieHH algorithm in \cite{zhu2020federated}, and learn heavy hitters in an interactive fashion by reducing the problem to a sequence of closed domain problems. 

Building on the observation that prefixes of heavy hitters will also be heavy, the algorithm discovers heavy hitters by building a trie (prefix tree) layer by layer.
Each layer of the trie stores a set of common prefixes of length corresponding to the depth of the layer. The trie is built in an iterative fashion starting from the root up to its maximum length. At each layer, we collect responses from a group of users, who only contribute by indicating one character after a common prefix from the previous layer. For example, if `COV' is a common prefix that the algorithm learned in the previous layer, and a user types a word `COVID-19', the user will only contribute their data by contributing `COVI' instead of the entire word `COVID-19'. Moreover, we apply randomization techniques with LDP guarantees on each user contribution. To bound the overall privacy loss for a user, we minimize user participation by restricting each user to contribute to at most one layer in the entire process and bound the number of items each user can contribute in their participation. The details of the algorithm are described in \cref{alg:ldp_triehh}.

\begin{algorithm}[ht!]
\caption{LDP TrieHH.}
\begin{algorithmic}[1]
\STATE \textbf{Input:} Set of characters $\cX$, end of word symbol $\eow \in \cX$, maximum tree depth $D$, list of known words $\cV$, sets of users $U$ and their local datasets $\{S_i\}_{i \in U}$. User contribution bound $B$. Per-layer number of users $N$. %
Per-layer maximum prefix $\eta_{\rm max}$.
A sampling function $\sample(S, B)$~that samples $B$ elements from a local dataset $S$.
\STATE \textbf{Output:} Set of heavy hitters $H$.
\STATE[] \coloredcomment{(Server) Intialization}
\STATE $H = \emptyset, P_0 = \cX$.
\FOR{$i = 1, 2, \ldots, D$}
\STATE[] \coloredcomment{(Server) Sample clients and broadcast learned prefixes.}
\STATE Sample a set of $N$ users $U_i$ that have not participated before.
\STATE Broadcast $P_{i-1}$ to users in $U_i$.
\FOR{$j \in U_i$}
\STATE[] \coloredcomment{(Local device) Local data selection and LDP randomization.}
\STATE Get a set of possible length-$i$ heavy prefixes.
\[
    \cZ_i = \{\bz x \mid \forall \bz \in P_{i-1}, x \in \cX \}.
\]
    \STATE Get a set of prefixes of length $i+1$ contributed by user $j$,
        \[
            S^{[:i+1]}_{j} = \{\bz[:i+1] \mid \bz \in S_{j}, \bz \notin \cV, \bz[:i+1]  \in \cZ_i,  \}.
        \]
    \STATE Sample a subset of $B$ elements
    $
        \tilde{S}^{[:i+1]}_{j} = \sample(S^{[:i+1]}_{j}, B).
    $
    \STATE Initialize a vote vector $\bv_j = \mathbf{0}_{|\cZ_i|}$
    \FOR{$\bz \in \tilde{S}^{[:i+1]}_{j}$}
        \STATE $\bv_j = \bv_j + \text{Multi-hot}(\mathbf{SS}(\bz, \cZ_i \cup \{\unk \}, \eps), \cZ_i)$
    \ENDFOR
    \STATE Send $\bV_j$ to the server.
    \ENDFOR
    \STATE[] \coloredcomment{(Server) Aggregation and obtain new heavy prefixes and words.}
    \STATE Aggregate all user votes $\bv^{U_i} = \sum_{j \in U_i} \bv_j.$
    \STATE Let $\tau_i$ be the value of the $\eta_{\rm max}$ largest entry in $\bv^{U_i}$.
\STATE $P_i = \{\bz \in \cZ_i \mid \bv_i[\bz] \ge \tau_i\}$
\STATE $H = H \cup \{\bz \in P_i \mid \bz[-1] = \bot\}$
\ENDFOR
\STATE \textbf{Return} $H$.
\end{algorithmic}
\label{alg:ldp_triehh}
\end{algorithm}

\paragraph{Local dataset sampler.} When a user's valid local contribution exceeds the contribution bound $B$, the user would perform local sampling to select a set of $B$ elements to report on, similar to prior work \cite{zhu2020federated, chadha2023differentially}. When sampling $B$ elements from a local dataset $S$, we consider two choices of the local sampler.
\begin{itemize}
    \item \gs. Return the top $B$ elements that are contributed the most.
    \item \rs. Return a uniform random subset of $B$ elements from the set of all contributed elements ignoring the number of times each element is contributed.
\end{itemize}
For both sampling methods, if the set $S$ contains less than $B$ distinct elements, we add copies of element $\unk$ %
to the set to make sure that the returned set is always of size $B$. %

\paragraph{Multiple-pass algorithm.} The algorithm can also be used as a building block for a multi-pass algorithm. The multi-pass algorithm performs builds multiple tries in a sequential order. Each trie is built by running \cref{alg:ldp_triehh}. After building each trie, we add the learned heavy hitters to the list of known words so that later runs of the algorithm will focus on learning new heavy hitters. Each run of \cref{alg:ldp_triehh} will use a disjoint set of users and hence the privacy guarantee will be the same as the single-pass version.%

\paragraph{Privacy implication of \cref{alg:ldp_triehh}.} The algorithm provides privacy guarantees in various aspects. First, instead of each user sending their entire contribution, we access user data only through queries that reveals limited information about their local contribution, in line of the \textit{data minimization} principle of federated analytics \cite{10.1145/3500240}.
This is done by capping each user's contribution to one contribution per day on average, as well as clipping each user contribution into prefixes.
Moreover, by deploying local randomization and central aggregation schemes, the algorithm offers differential privacy guarantees at different levels.
Below we describe the privacy guarantees of \cref{alg:ldp_triehh}. The theorems below follow from the privacy guarantee of \cref{alg:subset_selection} and Theorem 3.2 in \cite{feldman2023stronger}.
 \begin{theorem}\label{thm:ldp}
    With either \gs~or \rs~as the local sampler, \cref{alg:ldp_triehh} is $\eps$-LDP at item-level.
\end{theorem}

\begin{theorem}\label{thm:aggregated_privacy}
    With either \gs~or \rs~as the local sampler, the final output $H$, as well as the intermediate sets of heavy prefixes $P_i, i \in [D]$, obtained from \cref{alg:ldp_triehh} satisfy $(\eps', \delta)$-central DP at item-level, where $\eps'$ can be numerically computed using Theorem 3.2 in \cite{feldman2023stronger} with LDP parameter $\eps$ and number of contributions $\ns = NB$. For any $\delta \in [0, 1]$, and $\eps \le \ln\Paren{\frac{n}{8\ln(2/\delta)} - 1}$, $\eps'$ satisfies that 
    \[
        \eps' \le   \ln\Paren{1 + (e^\eps - 1)\Paren{\frac{4\sqrt{2\ln(4/\delta)}}{\sqrt{(e^\eps + 1)\ns}} + \frac4\ns}}.
    \]
\end{theorem}

\paragraph{Aggregation and access control on the server.} In practical implementation, this can be done via Secure Aggregtaion~\cite{bonawitz2017secure}, which limits the server side information to the aggregated votes through secure multi-party computation. Another way is to leave each individual vote vector ephemeral on the server and only store the aggregated result. By proper access control, an algorithm or analyst can only access the stored aggregated result, which ensures that the overall privacy guarantee of the heavy hitter algorithm can benefit from the amplification provided by aggregation.

\paragraph{Discussion on (approximate) $k$-anonymity.} $k$-anonymity is also an important privacy measure which requires that each of the released result in an algorithm has to be contributed by at least $k$ distinct users. In \cref{alg:ldp_triehh}, due to the added noise in the process, it is hard to guarantee an exact $k$-anonymity guarantee. Instead, we show that the algorithm satisfies an approximate version of the $k$-anonymity guarantee, defined below.
\begin{definition}[Approximate $k$-anonymity] \label{def:k-anon}
An algorithm $\cA$ that outputs a set of heavy hitters based on input set $\bS$ is said to satisfy $\alpha$-approximate $k$-anonymity, if $\bz \in \cZ$, which appeared less than $k$ times in $\bS$, we have
\[
    \probof{\bz \in \cA(\bS)} \le 1 - \alpha.
\]
\end{definition}
Then we can prove the following lemma on \cref{alg:ldp_triehh}.
\begin{lemma} \label{thm:k-anon}
Let $s = \eta_{\rm max} \cdot |\cX| + 1$,  $d = \ceil{\frac{s}{e^\eps + 1}}$, and $p = \frac{d e^\eps}{d e^\eps + s - d}, q = \frac{d - p}{s - 1}$. Let $\tau$ be the random variable obtained by getting the frequency of the $\eta_{\rm max}$-th most frequent element in a dataset constructed by adding $N\cdot B$ independent uniform subsets of $\{1, 2, \ldots, s-1\}$ with size $d-2$. Let $f(k) = f_1(k) + f_2(k)$ where $f_1(k)  \sim \text{Binomial}(k, p), f_2(k) \sim \text{Binomial}(N\cdot B - k, q)$ are two binomial random variables. Then outputs from \cref{alg:ldp_triehh} with length at least $\log_{|\cX|} \eta_{\rm max}$ satisfies $\alpha$-approximate $k$-anonymity if $\probof{f(k) \ge \tau} \le 1 - \alpha.$
\end{lemma}

\section{Case study of OOV discovery in Gboard inputs}
To study the effectiveness of the algorithm, we conduct experiments using the Gboard en-US population, to learn heavy hitters within a pre-specified list of target words, $\cZ_{\rm en}$. We keep $\cZ_{\rm en}$ within the existing known en-US vocabulary to avoid learning sensitive OOV words in undesirable parameter settings. To make the distribution of the words closer to the long-tailed distribution of OOV words, $\cZ_{\rm en}$ only contains words that are in the bottom 5\% percent of user daily contributions.

The main goal of the experiments is to identify parameter settings and algorithmic choices that would provide good performance-utility tradeoff in real OOV discovery tasks. Since the experiments are conducted on a large number of real user devices, we keep the number of experiments limited with the main purpose of obtaining engineering insights instead of providing an extensive scientific study.

We measure the quality of the discovered words by computing an estimated coverage of discovered words. To do this, we perform a counting task on $\cZ_{\rm en}$ and obtain the number of times that users contribute each word within a group of users. The estimated coverage of discovered words is computed by the total number of user contributions in the discovered set of words divided by the total number of user contributions in the entire target set.

In the experiments, we set $\cX$ to be \textsc{string.printable} in Python, which consists of single digits, upper and lower case letters, and other 38 special characters (100 characters in total). We fix $\eta_{\rm max} = 10000$ in light of system constraints. In our experiments, we observe that the coverage of the target words saturates after a depth of 15, hence we fix $D = 15$ in our explorations.

\begin{figure*}[h]
    \centering
    \includegraphics[width = 0.6 \textwidth]{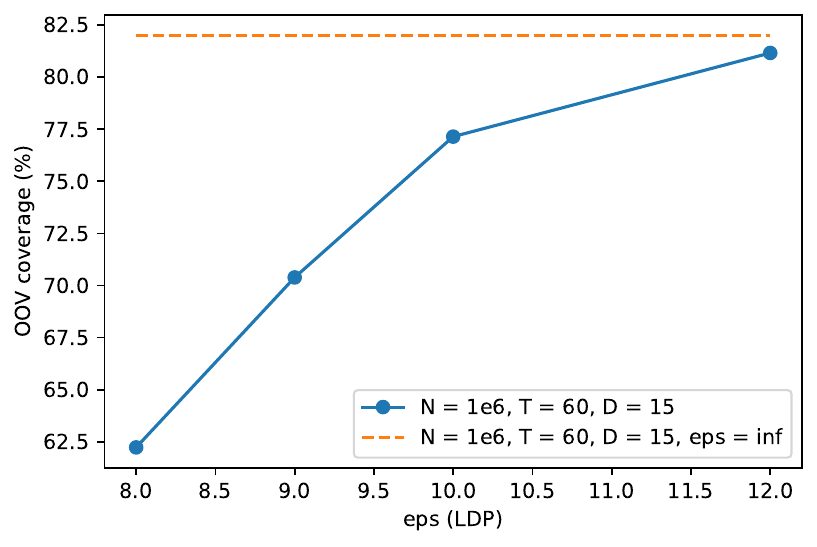}
    \caption{The estimated coverage of the recovered OOVs vs local privacy parameter $\eps$.}
\label{fig:varying_eps}
\end{figure*}

\paragraph{Privacy parameter.} We present results for the case when $N = 10^6$, $D= 15$, $T = 60$ with \gs~in \cref{fig:varying_eps}. As the figure shows, there is a tradeoff between the LDP parameter $\eps$ and the coverage of the discovered words. The dotted orange line marks the coverage we would get if no LDP is applied. We also observe that the decrease for coverage is relatively small for $\eps \ge 10$. 

\begin{table*}[h]
\centering
\caption{Coverage under different choices of local sampler and number of passes.  We fix $T = 60$, $D = 15$ and $\eps = 10$. For the one-pass algorithm, we use $N = 10^6$ and for the two-pass algorithm, we use $N = 5 \times 10^5$. Hence the total number of users is fixed to be $1.5 \times 10^7$. \text{GS} stands for \gs, and \text{RS} stands for \rs.}
\begin{tabular}{|c|c|c|c|}
\hline
 Setting & \text{GS} \& one-pass  & \textsc{RS} \& one-pass & \textsc{RS} \& two-pass \\ \hline
 Coverage & 77.1\% & 83.1\% & 92.1\% \\ \hline
\end{tabular}
\label{tab:comparison}
\end{table*}
\paragraph{Choice of the sampler and number of passes.} As shown in \cref{tab:comparison}, the choice of sampler and number of passes make a significant difference on the coverage of obtained set of heavy hitters. Switching from \gs~to \rs~alone increases the coverage of the discovered words by 6\%. The observation is similar to that of \cite{chadha2023differentially}, which shows that \rs~performs better than weighted sampling, where items are sampled based on their local frequencies. Conducting a two-pass algorithm further improves the coverage of discovered words by 9\%. Note here that we reduce the number of users per layer by half so the total number of users remains unchanged. \cite{chadha2023differentially} also observes that a multi-pass algorithm will improve the performance of trie-based heavy hitter algorithms. \cite{chadha2023differentially} lowers the privacy budget in each layer to keep the overall per-user privacy budget the same through privacy accounting while we lower the number of users per layer.

\subsection{OOV discovery in production population (en-US).} For OOV discovery in en-US production population, we choose $N = 5 \times 10^5, T = 60, \eps = 10, D = 15$. We use \gs~as the local sampler and perform two passes to build two tries. The total number of users involved in the process is $15$ million. For differential privacy guarantees, we set the privacy unit to be item-level and bound the number of contributions per users to 60. The user local dataset is collected in a span of 60 days, and hence each user contributes one word per day on average.
As discussed in \cref{thm:aggregated_privacy}, our LDP guarantee would imply a stronger central DP guarantee after aggregation at the server side. We obtain an item-level central DP guarantee of $\eps_{\rm central} = 0.315, \delta = 10^{-10}$ with the privacy amplification analysis and calculation tool provided in \cite{feldman2021hiding}. %
Moreover, plugging in the parameter setting in \cref{thm:k-anon}, \cref{alg:ldp_triehh} satisfies approximate $k$-anonymity guarantee with $k = 45$ and $\alpha = 0.95$. 

We test the coverage of the discovered words by computing the ratio between the number of OOV words in the discovered set and the total number of OOV words contributed by a test en-US population. Result shows that the algorithm allows us to discover words that account for 16.8\% of the OOV words. The discovered words include new trendy words such as \textit{afternoonish, dragonborn}, informal spelling of existing words such as \textit{Soooooo, srry, yayyy}, untypical capitalization such as \textit{athena, 
georgetown}, non-English words such as \textit{quede} (\textit{can} in Spanish), \textit{tiene} (\textit{has} in Spanish), and many others.

\section{Discussion}
We present an algorithm for federated discovery of OOV words from user inputs in Gboard with differential privacy guarantees. We demonstrate the effectiveness of the algorithm in  a production-level Gboard application with a large user population. The work opens up several directions for future exploration. (1) Develop better algorithms to further enhance the provided privacy guarantee, \eg lower the LDP parameter, or provide user-level differential privacy guarantee; (2) Explore the possibility of deploying the algorithm in languages with a smaller user population; (3) Provide externally verifiable privacy claims through technologies such as trusted execution environments (TEEs) ~\cite{tee_wiki}. Important progress on (3) has been made in the timely (and in-progress) work of \cite{eichner2024confidential} where they built a confidential and externally verifiable system for federated computations.  

\section{Acknowledgement}
The authors would like to thank Brendan McMahan, Micheal Riley, and Yuanbo Zhang for their feedback on the work, and Zheng Xu for their comments on the draft.

\bibliography{references}
\bibliographystyle{unsrt}
\appendix
\input{proofs}
\end{document}

%% file: glodef.tex
\usepackage{lmodern}
\usepackage{xspace}
\usepackage[protrusion=true,expansion=true]{microtype}
\usepackage{amsfonts,amsmath,amssymb,amsthm}

\usepackage[colorlinks,citecolor=blue,bookmarks=true]{hyperref}

\usepackage{multirow}
\usepackage{dsfont} %
\usepackage{fullpage}

\usepackage{algorithmic, algorithm}

\usepackage[shortlabels]{enumitem}
\setitemize{noitemsep,topsep=0pt,parsep=0pt,partopsep=0pt}
\setenumerate{noitemsep,topsep=0pt,parsep=0pt,partopsep=0pt}

\makeatletter
\@ifundefined{theorem}{%
  \theoremstyle{definition}
  \newtheorem{definition}{Definition}
  \theoremstyle{plain}
  \newtheorem{theorem}{Theorem}
  
  \newtheorem{lemma}{Lemma}

  \theoremstyle{remark}

}{}
\makeatother

\newcommand{\ignore}[1]{}

\newcommand{\RR}{\mathbb{R}}

\def \bS     {{\bf S}}

\def \cA     {{\cal A}}

\def \cO     {{\cal O}}

\def \cV     {{\cal V}}

\def \cX     {{\cal X}}
\def \cY     {{\cal Y}}
\def \cZ     {{\cal Z}}

\newcommand{\eg}{\textit{e.g.,}}%

\def \ceil#1{{\lceil{#1}\rceil}}

\def \Paren#1{{\left({#1}\right)}}

\newcommand{\eqdef}{{:=}}

\newcommand{\probof}[1]{\Pr\Paren{#1}}

\def \iff    {{\it iff }}
\def \th     {{\rm th }}

\def\ignore#1{}

\newcommand{\bi}{\begin{itemize}}
\newcommand{\ei}{\end{itemize}}

\def\orpro{\mathop{\mathchoice
   {\vee\kern-.49em\raise.7ex\hbox{$\cdot$}\kern.4em}
   {\vee\kern-.45em\raise.63ex\hbox{$\cdot$}\kern.2em}
   {\vee\kern-.4em\raise.3ex\hbox{$\cdot$}\kern.1em}
   {\vee\kern-.35em\raise2.2ex\hbox{$\cdot$}\kern.1em}}\limits}

\def\andpro{\mathop{\mathchoice
 {\wedge\kern-.46em\lower.69ex\hbox{$\cdot$}\kern.3em}
 {\wedge\kern-.46em\lower.58ex\hbox{$\cdot$}\kern.25em}
 {\wedge\kern-.38em\lower.5ex\hbox{$\cdot$}\kern.1em}
 {\wedge\kern-.3em\lower.5ex\hbox{$\cdot$}\kern.1em}}\limits}

\def\simge{\mathrel{%
   \rlap{\raise 0.511ex \hbox{$>$}}{\lower 0.511ex \hbox{$\sim$}}}}

\def\simle{\mathrel{
   \rlap{\raise 0.511ex \hbox{$<$}}{\lower 0.511ex \hbox{$\sim$}}}}

%% file: locdef.tex
\renewcommand{\algorithmicrequire}{\textbf{Input:}}

\newcommand{\ns}{n}

\newcommand{\eps}{\varepsilon}

%% file: proofs.tex
\section{Missing analysis.}
\paragraph{Proof of \cref{thm:k-anon}}
We focus on the voting process in one layer and prove that after the $(\ceil{\log_{|\cX|} \eta_{\rm max}} - 1)$th layer, a heavy prefix will be kept with probability at most $1 - \alpha$ if it is contributed by less than $k$ users if $\probof{f(k) \ge \tau} \le 1 - \alpha.$, which would imply \cref{thm:k-anon}.

At layer $i$, users are voting on the set $\cZ_i \cup \{ \eow \}$. After the $(\ceil{\log_{|\cX|} \eta_{\rm max}} - 1)$th layer, $|\cZ_i|  = |P_i|\times |\cX| =   \eta_{\rm max} \cdot |\cX|$, and the size of the set is $s = |\cZ_i \cup \{ \eow \}| = \eta_{\rm max} \cdot |\cX| + 1$. Without loss of generality, we simply assume there is a natural numbering on the set and users contribute elements in the set $[s]$. In this phase, users contribute $n = NB$ items in total. Note that with either \gs~or \rs, each user contributes one item at most once. Hence it would be enough to show that a heavy prefix will be kept with probability at most $1 - \alpha$ if it is contributed less than $k$ times. Hence from now on, we will ignore how items are contributed by different users, and simply denote the set of contributed elements by $\{z_1, \ldots z_\ns\}$. We will focus on the item numbered $1$, and the rest of the items will follow by the same argument.

Let $\bv^U \in \RR^{s}$ be the aggregated vote vector. $\forall j \in [\ns]$, let $\bv_j$ be the vote vector obtained by randomizing $z$ (running \cref{alg:subset_selection}). Then we have
\[
    \bv^U = \sum_j \bv_j.
\]

Let $\tau_{\bv}$ be the $\eta_{\rm max}$-th largest element in $\bv^U$, item 1 will only be kept if $\bv^U[1] \ge \tau_{\bv}$.  Next we construct a set of vote vectors $\bv'_j$ such that there exists a coupling between each pair of $\bv'_j$ and $\bv_j$ satisfying $\bv'_j[1] = \bv_j[1]$ and $\bv'_j[t] \le \bv_j[t]$ for all $t \neq 1$ with probability one. Then setting $\bv^{U'} =  \sum_j \bv'_j$, we will have that $\bv^{U'}[1] = \bv^{U}[1]$ and $\bv^{U'}_j[t] \le \bv^{U}[t]$ for all $t \neq 1$ with probability one. Let $\tau'_{\bv}$ be the $\eta_{\rm max}$-th largest element in $\bv^{U'}$. Then we have
\[
    \probof{\bv^U[1] \ge \tau_{\bv}} \le \probof{\bv^{U'}[1] \ge \tau'_{\bv}},
\]
which would provide an upper bound on the probability that a prefix will be kept.

We state the construction of $\bv'_j$ below. Let $d = \ceil{\frac{s}{e^\eps + 1}}$. 
\begin{itemize}
    \item Case 1: $z_j = 1$. Set $\bv_j'[1] = 1$ with probability $p = \frac{d e^\eps}{d e^\eps + s - d}$ as defined in \cref{alg:subset_selection}. Draw a random subset with size $d - 2$ from $[s] \setminus \{1\}$ and set the coordinates within the subsets to 1. Set the rest of coordinates to 0.
    \item Case 2: $z_j \neq 1$. Set $\bv_j'[1] = 1$ with probability $q = \frac{d - p}{s - 1}$. Draw a random subset with size $d - 2$ from $[s] \setminus \{1\}$ and set the coordinates within the subsets to 1. Set the rest of coordinates to 0.
\end{itemize}
Next we describe the coupling by presenting a random mapping from $\bv_j'$ to $\bv_j$.
\begin{itemize}
    \item Case 1: $z_j = 1$. Add uniformly random $d - \|\bv_j'\|_1$ zero coordinates of $\bv_j'$ in $[s] \setminus \{1\}$ and set them to 1. Let $\bv_j$ be the resulting vector.
    \item Case 2: $z_j \neq 1$. If $\bv_j'[z_j] = 0$, let $p' = \frac{d-2}{s-1}$, change $\bv_j'[z_j]$ to 1 with probability $\frac{p - p'}{1-p'}$. Add uniformly random $d - \|\bv_j'\|_1$ zero coordinates of $\bv_j'$ in $[s] \setminus \{1\}$ and set them to 1. Let $\bv_j$ be the resulting vector.
\end{itemize}

It is clear from the construction that the coupling satisfies monotonicity on each of the coordinates. It can be shown that the resulting vector $\bv_j$ satisfies the output distribution from \cref{alg:subset_selection}. Let $n_1$ be the number of times that 1 appears in $\{z_1, \ldots z_\ns\}$. We have that $\bv^{U'}[1]$ is the sum of two binomial random variables $\text{Binomial}(n_1, p)$, and $\text{Binomial}(N\cdot B - n_1, q)$. Moreover, $\tau_\bv'$ is at least as large as the $\eta_{\rm max}$ largest coordinate in coordinates $[s] \setminus \{1\}$ in $\bv_j'$. Hence we conclude the proof of the Lemma.